\documentclass[11pt,a4paper]{article}
\usepackage[T1]{fontenc}
\usepackage[utf8]{inputenc}
\usepackage{lmodern}
\usepackage{geometry}
\usepackage{amsmath,amssymb,bm,mathrsfs}
\usepackage{booktabs,longtable,graphicx,caption,float}
\usepackage[authoryear,round]{natbib}
\usepackage{microtype}
\usepackage{xurl}
\usepackage[colorlinks=true,allcolors=blue!55!black]{hyperref}
\usepackage{xcolor}
\newcounter{scheme}

\providecommand{\urlprefix}{}
\newcommand{\doi}[1]{\href{https://doi.org/#1}{doi:#1}}
\newcommand{\beq}{\begin{equation}}
\newcommand{\eeq}{\end{equation}}

\title{Condensation-related expansion work of water vapor as a major pathway for kinetic energy generation in the lower atmosphere}
\author{Anastassia M. Makarieva$^{1,2,3}$ \and Andrei V. Nefiodov$^{1,2,3}$\\[0.5em]
\small $^1$Petersburg Nuclear Physics Institute, NRC Kurchatov Institute, Russia\\
\small $^2$Institute for Advanced Study, Technical University of Munich, Germany\\
\small $^3$Biotic Pump Greening Institute, S\~ao Jos\'e dos Campos, Brazil}
\date{}
\begin{document}
\maketitle

\begin{abstract}
Earth's atmosphere operates a steam cycle: water evaporates
from the surface, expands as vapor, condenses in colder air,
and returns as precipitation. Yet its power and
relationship to atmospheric mechanical power remain poorly
quantified. Applying the Clausius--Clapeyron relation between
saturated vapor expansion work and latent heat converted at
Carnot efficiency to 
vertically distributed condensation, we derive the steam-engine power as a sum of
the gravitational and condensation-related contributions,
$W_v=W_g+W_c$; it is not the net power of a complete moist-air
cycle.

Using twelve years of GPM/DPR spectral latent-heating profiles
and MERRA-2 fields, we estimate
$W_v\simeq4.1$~W~m$^{-2}$,  with
$W_c\simeq2.8$~W~m$^{-2}$. Separately, we diagnose total
atmospheric mechanical power $W=W_P+W_K\simeq4.3$~W~m$^{-2}$,
comprising gravitational power of precipitation
$W_P\simeq1.1$~W~m$^{-2}$ and kinetic energy generation
$W_K\simeq3.2$~W~m$^{-2}$, including
$2.4$~W~m$^{-2}$ below 500~hPa. Both correspondences
$W_v\simeq W$ and $W_c\simeq W_K$ lie within 20--30\%
uncertainty estimates.

To explain these correspondences, we propose that $W_c$
generates the potential energy associated with surface pressure
gradients. Cross-isobaric flow in the frictional lower atmosphere converts this energy into kinetic energy.
In a horizontally isothermal atmosphere, this positive pressure
work would be nearly cancelled by compression aloft.
An idealized circulation illustrates how differential heating can offset compression, 
allowing nearly isobaric upper return flow.
In the proposed mechanism, condensation-driven
circulation unlocks this thermal compensation thus providing a
major pathway for lower-atmospheric kinetic energy generation.
We discuss how changes in the coupling between condensation
and differential heating could test this proposed causal
relationship.
\end{abstract}

\section{Introduction}
\label{sec:intro}

Understanding the causes of changes in atmospheric circulation and their climatic consequences remains a major challenge in climate science \citep{shepherd2014,shaw2024}. An important part of this problem is identifying the physical mechanisms of atmospheric kinetic energy generation, particularly those associated with water vapor phase transitions. The global mean latent heat flux is approximately $80~\mathrm{W\,m^{-2}}$, nearly four times the sensible heat flux \citep{forster2021}. While these energy fluxes are relatively well quantified, establishing a robust quantitative relationship between water vapor phase transitions and atmospheric kinetic energy generation remains a challenge.

Currently, the magnitude of atmospheric mechanical power remains uncertain. Among more recent estimates, \citet{pa12} stated that the total power associated with atmospheric motions is not known precisely but is likely on the order of $5~\mathrm{W\,m^{-2}}$. \citet{pauluis15} cited the results of \citet{laliberte15} as supporting this magnitude. In contrast, \citet{bohren2023} reported a broader range of $2$--$5~\mathrm{W\,m^{-2}}$, whereas \citet{kleidon2021} argued, on theoretical grounds, that the rate should not exceed $2~\mathrm{W\,m^{-2}}$. Finally, observational estimates of Lorenz available potential energy generation yielded only about $1.2~\mathrm{W\,m^{-2}}$ \citep{romanski2013}. Part of this discrepancy concerns the distinction between kinetic energy generation by atmospheric winds, $W_K$, and total atmospheric power, $W=W_K+W_P$, which also includes the gravitational power required to lift precipitating water, $W_P$, of approximately $1~\mathrm{W\,m^{-2}}$ \citep{pa12}. Yet a robust observational assessment of both contributions using independent diagnostics remains lacking, thereby preventing a reliable validation of theoretical approaches against observations.

Theoretical predictions of atmospheric power generation are also subject to considerable uncertainty. In moist heat-engine formulations, mechanical work is commonly evaluated using entropy budgets that depend on irreversible processes associated with water vapor diffusion and cloud microphysics \citep{pauluis2002b,singh2022}. These processes are difficult to constrain independently, complicating the observational validation of theoretical power estimates \citep{jansen2026}. This highlights the need for transparent, quantitative predictions that can be tested against independent observations.

An important finding of moist atmospheric heat-engine theory
is that moisture-related irreversible processes can
substantially reduce thermodynamic efficiency below the
Carnot limit \citep{pauluis2002b,pa11}. However, a lower efficiency does not necessarily imply lower mechanical power. The maximum Carnot efficiency corresponds to the reversible limit of an infinitely slow cycle, while the power generated by an actual circulation depends on both its efficiency and the rate of energy transfer. Furthermore, latent and sensible heat fluxes are not independent of atmospheric circulation. They depend on winds and atmospheric temperature and humidity distributions, and could be substantially different in the absence of the water cycle. Thus, understanding the role of water vapor phase transitions in atmospheric power generation requires us to consider not only how moisture affects the efficiency of a given thermodynamic cycle, but also what circulations become possible when the water cycle is present.

A different perspective on the role of moisture in atmospheric power generation dates back to Heinrich Hertz, who in 1885 proposed that the atmosphere operates as a steam engine, with evaporation at the Earth's surface and condensation in clouds corresponding to the boiler and condenser, respectively \citep{mulligan1997}. Hertz estimated the mean temperature difference between evaporation and condensation to be approximately $15~\mathrm{K}$. At a temperature of about $300~\mathrm{K}$, this corresponds to a Carnot efficiency of approximately $5\%$. Applying this efficiency to the present-day global latent heat flux of about $80~\mathrm{W\,m^{-2}}$ yields a steam-engine power of approximately $4~\mathrm{W\,m^{-2}}$, remarkably close to the estimates of total atmospheric mechanical power discussed above. In view of the robust finding that the water cycle reduces thermodynamic efficiency relative to the Carnot limit, an important question is what underlying physical relationship, if any, could account for this numerical correspondence. In this paper, we assess this numerical correspondence using modern observations and investigate its possible physical basis.

A possible physical mechanism linking the atmospheric water cycle to kinetic energy generation has been proposed within condensation-induced atmospheric dynamics (CIAD), developed since 2007 \citep[for a review, see][]{makarieva19}. Condensation removes water vapor from the gas phase, introducing a mass sink into the gaseous continuity equation, while precipitation removes condensed water from the atmospheric column. Together with evaporation and atmospheric transport, these processes can maintain horizontal differences in atmospheric column mass and surface pressure, thereby contributing to kinetic energy generation in the lower atmosphere.

However, the relationship between condensation rate and
power generation, while yielding meaningful predictions
across diverse meteorological contexts, has remained
largely heuristic. Its connection to the thermodynamic
expansion work of the atmospheric steam cycle has not
been explicitly established. This raises a broader
question: does the water cycle determine atmospheric
circulation, or does it primarily modify the efficiency
of circulation driven by differential heating?

To address these questions, Section~\ref{sec:steam}
derives atmospheric steam-engine power from the
Clausius--Clapeyron law, separates its gravitational
and condensation-related components, and defines
the corresponding contributions to atmospheric
mechanical power.
Section~\ref{sec:methods} describes the observational data and diagnostic methods.
Section~\ref{sec:results} presents the global
power estimates and their uncertainties, and
compares steam-engine and atmospheric mechanical
power.
In Section~\ref{sec:discussion}, we examine the
relationship to condensation-induced atmospheric
dynamics, investigate how differential heating
can prevent the cancellation of lower-level
pressure work by compression aloft, and discuss
thermodynamic efficiency, causality, and implications
for predicting atmospheric circulation.
Appendices~\ref{app:WP} and~\ref{app:WK} provide
supporting derivations and computational details,
while Appendix~\ref{app:Lorenz} develops a simplified
mass-related Lorenz-type available-energy framework
to examine the relationship between surface pressure
gradients and differential heating.

\section{Water-vapor expansion and atmospheric mechanical power}
\label{sec:steam}

\subsection{Clausius–Clapeyron work and water-vapor expansion power}
\label{sec:CCTc}

The Clausius--Clapeyron relation for the saturated water-vapor pressure
$p_v^*$ is usually written as
\begin{equation}\label{cc}
\frac{dp_v^*}{dT}=\frac{\mathcal L p_v^*}{RT^2},
\end{equation}
or, equivalently,
\begin{equation}\label{cc_reform}
\frac{RT}{p_v^*}\,dp_v^*=\mathcal L\frac{dT}{T},
\end{equation}
where $R$ is the universal gas constant and $\mathcal L$ is the molar
latent heat of vaporization. The corresponding mass-specific latent
heat is $L=\mathcal L/M_v$, where $M_v$ is the molar mass of water.
In this form, the equation relates the expansion work of saturated
water vapor to latent heat converted at Carnot efficiency.

The Clausius--Clapeyron equation is commonly derived from the equality
of Gibbs free energies, or chemical potentials, of liquid water and
vapor \citep{andrews2010,bohren2023}. An alternative classical
derivation considers an infinitesimal reversible Carnot cycle
\citep{brunt34,landau1967,wallace2006}, whose net work is
$\Delta V\,\Delta p_v^*$. Neglecting the liquid-water volume and assuming
ideal-vapor behavior, $\Delta V\simeq RT/p_v^*$. Equating this work
to $\mathcal L\,\Delta T/T$ gives Eq.~\eqref{cc_reform}.

Equation~\eqref{cc_reform} provides the starting point for evaluating
the expansion work of saturated water vapor in the atmosphere. Since condensation
occurs over a range of temperatures and along many trajectories, we
weight the work by the amount of water condensing at each temperature.
Let $C\geq0$ and $E\geq0$ denote the local rates of net condensation
and net evaporation of water vapor, respectively (kg\,m$^{-3}$\,s$^{-1}$).
These rates represent opposite signs of the local net phase change:
at any given point and instant, at most one of them is
nonzero\footnote{Thus defined, $C$ and $E$ are scale-dependent.
If the entire Earth were represented by a single grid cell,
both would be zero in steady state. An appropriate spatial
scale should therefore be chosen to capture the separation
of condensation and evaporation occurring at different
environmental temperatures.}.
Consider the atmospheric volume $\Omega$ above a reference surface
area $\mathcal S$. For any quantity $X$ associated with a
condensation event, the condensation-weighted mean is defined as
\begin{equation}\label{cond_weight}
\langle X\rangle_c\equiv
\frac{\displaystyle\int_\Omega\overline{CX}\,d\Omega}
{\displaystyle\int_\Omega\overline C\,d\Omega}.
\end{equation}
\begin{equation}\label{Pc}
P_c\equiv\frac{1}{\mathcal S}
\int_\Omega\overline C\,d\Omega.
\end{equation}
Here an overbar denotes time averaging over the period considered.
The condensation rate per unit surface area $P_c$ need not equal the
surface precipitation $P$.

The water vapor condensing at temperature $T$ has previously
evaporated elsewhere and, before condensing, reached saturation
at temperature $T_1$. We refer to $T_1$ as the initial
saturation temperature associated with this condensation
contribution. Under saturated conditions, vapor pressure is
uniquely determined by temperature. Consequently, the expansion
work between the initial saturated state and the condensation
state calculated from the Clausius–Clapeyron relation depends only on $T_1$ and $T$, irrespective of the
particular atmospheric trajectory connecting these states.

For each condensation contribution, integrating
Eq.~\eqref{cc_reform} from $T$ to $T_1$ gives the
expansion work per mole of water vapor as the
saturation pressure decreases from its initial
value at $T_1$ to that at $T$.
Multiplying this work by the local molar
condensation rate $C/M_v$, averaging over time,
and integrating over the atmospheric volume,
we obtain the water-vapor expansion power
per unit surface area:
\begin{equation}\label{Wv}
\begin{aligned}
W_v
&=\frac{1}{\mathcal S M_v}
\int_\Omega
\overline{
C\int_T^{T_1}
\mathcal L(T')\frac{dT'}{T'}
}\,d\Omega\\
&=\frac{P_c}{M_v}
\left\langle
\int_T^{T_1}
\mathcal L(T')\frac{dT'}{T'}
\right\rangle_c\\
&\simeq LP_c
\left\langle\ln\frac{T_1}{T}\right\rangle_c,
\end{aligned}
\end{equation}
where the final approximation neglects the temperature
dependence of $\mathcal L$. 

Equation~\eqref{Wv} relates water-vapor expansion power to the condensation rate and the temperatures of the initial saturated and condensation states. The distribution of condensation temperatures can be estimated from GPM/DPR latent-heating profiles combined with atmospheric temperature data, while the initial saturation temperatures can be estimated from lifting-condensation levels derived from MERRA-2. These estimates provide the basis for calculating $W_v$ from observations, as described in the Methods section.

\subsection{Separation of gravitational and condensation-related expansion work}

We now separate water-vapor expansion work into gravitational and condensation-related components. 
Writing $\gamma\equiv p_v^*/p_d$, where $p_d$ is dry-air partial
pressure, gives the identity
\begin{equation}\label{dg}
d\ln p_v^*=d\ln p_d+d\ln\gamma.
\end{equation}
Thus the decrease of saturation vapor pressure can be separated into
changes of dry-air pressure and of the vapor-to-dry-air pressure ratio.

Hydrostatic equilibrium relates the vertical variation of air
pressure to height:
\begin{equation}\label{he}
RT\frac{\partial\ln p}{\partial z}=-Mg,
\end{equation}
where $M$ is the mean molar mass of gaseous air, taken approximately
constant, and $g$ is gravitational acceleration. To accuracy $O(\gamma)$, $\partial_z\ln p_d\simeq
\partial_z\ln p$. 

Let $z$ and $z_1$ denote the heights of the condensation and initial saturated states, respectively.
Evaluating the vertical dry-air pressure difference in the
local hydrostatic column for each condensation event gives
its gravitational contribution to the expansion power:
\begin{equation}\label{Wg}
\begin{aligned}
W_g&\equiv-\frac{RP_c}{M_v}\left\langle
\int_{z_1}^{z}T(z')
\frac{\partial\ln p_d}{\partial z'}\,dz'
\right\rangle_c\\
&\simeq\frac{M}{M_v}gP_c(z_c-\langle z_1\rangle_c),
\end{aligned}
\end{equation}
where $z_c\equiv\langle z\rangle_c$ is the mean height of condensation.
While $W_v$ depends on condensation temperatures,
$W_g$ depends on condensation heights, which we also
estimate using GPM/DPR and MERRA-2 data.

The remaining condensation-related power is
\begin{equation}\label{Wc}
\begin{aligned}
W_c&\equiv W_v-W_g\\
&\simeq LP_c\left\langle\ln\frac{T_1}{T}\right\rangle_c
-\frac{M}{M_v}gP_c(z_c-\langle z_1\rangle_c).
\end{aligned}
\end{equation}
Thus, the water-vapor expansion power decomposes as
\begin{equation}\label{Wv_split}
W_v=W_g+W_c.
\end{equation}

To clarify the physical meaning of $W_c$,
we use Eqs.~\eqref{cc_reform} and \eqref{dg} to write the
expansion work per mole associated with each condensation
contribution as
\begin{equation}
\int_T^{T_1}\mathcal L(T')\frac{dT'}{T'}
=
-R\int_{p_{d}(z_1)}^{p_d} T\,d\ln p_d'
-R\int_{\gamma_1}^{\gamma} T\,d\ln\gamma',
\end{equation}
where the integrals on the right are taken between the
initial saturated state and the condensation state.
The vertical hydrostatic contribution of the first integral,
after condensation weighting and multiplication by $P_c/M_v$,
gives $W_g$ in Eq.~\eqref{Wg}.
For the idealized calculation below, we neglect
horizontal and temporal changes in dry-air pressure.
The remaining condensation-related power is then
\begin{equation}\label{Wc_gamma}
W_c\simeq
-\frac{RP_c}{M_v}\left\langle
\int_{\gamma_1}^{\gamma}
T(\gamma')\,d\ln\gamma'
\right\rangle_c.
\end{equation}

Consider an idealized saturated trajectory with a fixed
amount of dry air, where $\gamma$ decreases from
$\gamma_1$ to $\gamma_2$ solely through condensation.
Since $\gamma$ equals the molar ratio of water vapor
to dry air, the amount condensed is proportional to
$-d\gamma$. Approximating the temperature in the
preceding integral by $T_c \equiv \langle T \rangle_c$, we obtain
\begin{equation}\label{Wc_kappa}
\begin{aligned}
W_c&\simeq\frac{RP_cT_c}{M_v(\gamma_1-\gamma_2)}
\int_{\gamma_2}^{\gamma_1}
\ln\frac{\gamma_1}{\gamma}\,d\gamma\\
&=\frac{RP_cT_c}{M_v}(1+\kappa),\\
\kappa&\equiv
\frac{\gamma_2/\gamma_1}{1-\gamma_2/\gamma_1}
\ln\frac{\gamma_2}{\gamma_1}\leq0.
\end{aligned}
\end{equation}

The ratio $\gamma_2/\gamma_1$ represents the fraction of
water vapor remaining uncondensed\footnote{For a hypothetical isothermal vapor atmosphere,
$p_d(\gamma_1-\gamma_2)$ represents the amount of vapor
condensed per unit volume, expressed in pressure units.
Multiplying this quantity by $1+\kappa$ gives
$p\ln(p/p_0)+p_0-p$, where $p=\gamma_2p_d$ and
$p_0=\gamma_1p_d$, corresponding to the potential
energy per unit volume in Eq.~(I$a'$) of
\citet{Margules1901}.}. For
$\gamma_2/\gamma_1=0.3$, $\kappa\simeq-0.52$,
while $\kappa\to0$ for complete condensation.
In the opposite limit, $\gamma_2/\gamma_1\to1$, condensation
vanishes, $\kappa\to-1$, and $W_c\to0$. With no
condensation, the condensation-weighted mean quantities
become undefined. Water vapor then behaves as a
noncondensable gas, and separating its expansion work from
that of the other atmospheric gases has no physical meaning.

For the atmosphere as a whole, an effective value of
$\kappa$ can be estimated from the condensation-related
power $W_c$, the condensation rate per unit surface area $P_c$, and the
mean condensation temperature $T_c$:
\begin{equation}\label{kappa_eff}
\kappa_{\rm eff}
=\frac{M_v W_c}{R P_c T_c}-1
=\frac{M_v(W_v-W_g)}{R P_c T_c}-1.
\end{equation}

\subsection{Mechanical atmospheric power: $W=W_P+W_K$}
\label{sec:mechanical}
Atmospheric pressure work per unit surface area is
\begin{equation}\label{W}
W=\frac{1}{\mathcal S}\int_\Omega
\overline{p\nabla\cdot\mathbf v}\,d\Omega
=-\frac{1}{\mathcal S}\int_\Omega
\overline{\mathbf v\cdot\nabla p}\,d\Omega,
\end{equation}
where $\mathbf v=\mathbf u+w\mathbf e_z$ is gaseous-air velocity,
$\mathcal S$ is the reference surface area, and the boundary flux of
$p\mathbf v$ is taken to vanish. Separating the vertical and horizontal contributions
to pressure work and using hydrostatic equilibrium,
we obtain
\begin{equation}\label{Wsum}
W=W_P+W_K,
\end{equation}
where
\begin{equation}\label{WP}
W_P=gP_c(z_c-z_e)
\end{equation}
is the gravitational contribution to pressure work, and
\begin{equation}\label{WK}
W_K=-\frac{1}{\mathcal S}\int_\Omega
\overline{\mathbf u\cdot\nabla_h p}\,d\Omega
\end{equation}
is kinetic energy generation by horizontal pressure gradients.
Here $P_c$, defined in Eq.~\eqref{Pc}, is the
condensation rate per unit surface area, $z_c$ and $z_e$ are the corresponding condensation- and evaporation-weighted
heights (Appendix~\ref{app:WP}). 

In pressure coordinates, the evaluation of total
atmospheric pressure work $W$ leads to the so-called
$\alpha\omega$ formulation, where
$\alpha\equiv1/\rho$ is specific volume and
$\omega\equiv Dp/Dt$ is pressure vertical velocity.
This formulation involves cancellation between large
positive and negative contributions, so even small
numerical errors in $\omega$ or mass continuity may
strongly affect the global residual
\citep{steinheimer2008,wang2024global}.
Here we diagnose $W_K$ directly from horizontal pressure
gradients and calculate $W_P$ from the height-weighted
local net condensation and evaporation rates $C$ and $E$.

The gravitational contribution $W_g$ to water-vapor expansion is
distinct from $W_P$. From Eqs.~\eqref{Wg} and \eqref{WP},
\begin{equation}\label{Wg_WP_ratio}
\frac{W_g}{W_P}\simeq
\frac{M}{M_v}\frac{z_c-\langle z_1 \rangle_c}{z_c-z_e}.
\end{equation}
This ratio reduces to $M/M_v\simeq1.6$ when
$\langle z_1 \rangle_c=z_e$, i.e. when water vapor enters saturation at the same mean height where it evaporated,
which is not generally the case.

\section{Data and methods}
\label{sec:methods}

We estimate $W_v$, $W_g$, and $W_P$ from the GPM Dual-frequency
Precipitation Radar (DPR) spectral latent-heating product
(SLH V07; \citealp{hamada2025slh}), together with
contemporaneous three-hourly MERRA-2 pressure-level data
\citep{gelaro2017merra2}, and obtain $W_c=W_v-W_g$.
The analysis covers the entire period of availability of SLH V07,
from 8 March 2014 to 28 February 2026, comprising twelve
March--February periods.

The GPM analysis covers $65^\circ$S--$65^\circ$N.
To account for unequal latitude sampling, we first
average within $5^\circ$ latitude bands and then
weight the results by the area of each band relative
to the entire Earth's surface area. Thus, the
resulting fluxes and powers are normalized by the
total Earth's area, while contributions poleward
of $65^\circ$ are not included.

The GPM/DPR radar scans swaths approximately 250--280~km wide, with
49 footprints per cross-track scan. Each footprint is about 5~km across
(approximately 25~km$^2$), and SLH provides temperature tendencies
$\dot T$ in 80 layers of 250~m thickness. For each footprint, we take
MERRA-2 temperature, pressure, humidity, and density from the nearest
grid point, interpolating between consecutive three-hourly records and
to the SLH height levels. We estimate the local rates of net condensation
$C$ and atmospheric evaporation $E_a$ as
\begin{equation}\label{eq:gpm-CE-proxy}
C\simeq\frac{\rho c_p}{L}\max(\dot T,0),\qquad
E_a\simeq\frac{\rho c_p}{L}\max(-\dot T,0),
\end{equation}
where $\dot T$ is in K~s$^{-1}$, $\rho$ is MERRA-2 air density,
$c_p=1004$~J~kg$^{-1}$~K$^{-1}$, and
$L=2.5\times10^6$~J~kg$^{-1}$. We approximate the retrieved heating
by vapor--liquid phase change, without separately removing contributions
from freezing and melting.

\begin{figure}[H]
\centering
\includegraphics[width=0.85\linewidth]{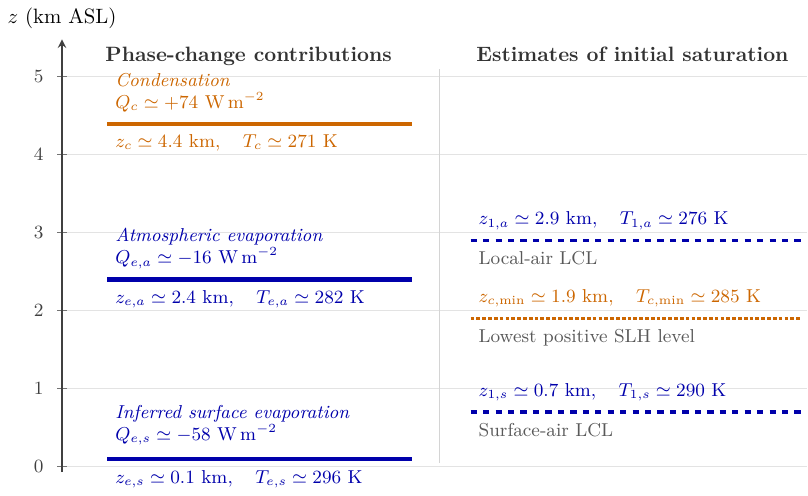}
\caption{Vertical arrangement of condensation, evaporation,
and estimates of initial saturation. The dashed lines indicate the lifting
condensation levels of surface air and air at atmospheric evaporation
levels, used in the two-LCL estimate.
The dotted line indicates $z_{c,\min}$, the
positive-heating-weighted mean of the lowest levels
of positive SLH heating in individual footprints,
used in the one-onset estimate.}

\label{fig:twoLCL_scheme}

\end{figure}

Surface evaporation is not retrieved by SLH and is inferred from
steady-state mass balance. With $P_{e,a}$ and $P_{e,s}$ denoting
atmospheric and surface evaporation rates per unit area, we define
the signed latent-heating contributions
\begin{equation}\label{eq:gpm-Qs}
\begin{aligned}
Q_c&\equiv LP_c,\\
Q_{e,a}&\equiv-LP_{e,a},\\
Q_{e,s}&\equiv-LP_{e,s}=-(Q_c+Q_{e,a}),
\end{aligned}
\end{equation}
using $P_c=P_{e,a}+P_{e,s}$. To account for unequal latitude sampling,
we average within $5^\circ$ latitude bands and then weight by band
area over $65^\circ$S--$65^\circ$N. 
The twelve-period Earth-area-normalized estimates are
$Q_c\simeq74$, $Q_{e,a}\simeq-16$, and
$Q_{e,s}\simeq-58$~W~m$^{-2}$
(Fig.~\ref{fig:twoLCL_scheme}).

The heights used to calculate $W_P$ are expressed above mean
sea level (ASL), as in Fig.~\ref{fig:twoLCL_scheme}.
The corresponding mean heights are
$z_c\simeq4.44$~km for condensation,
$z_{e,a}\simeq2.35$~km for atmospheric evaporation, and
$z_{e,s}\simeq0.13$~km for inferred surface evaporation, estimated
from the mean local surface elevation. Thus,
\begin{equation}
z_e=\frac{P_{e,a}z_{e,a}+P_{e,s}z_{e,s}}{P_c}
\simeq0.61~\mathrm{km}.
\end{equation}
Equation~\eqref{WP} then gives
$W_P=gP_c(z_c-z_e)\simeq1.11$~W~m$^{-2}$.
Setting $z_{e,s}=0$ instead increases $W_P$ to
$1.14$~W~m$^{-2}$, a difference of approximately 3\%.

We evaluate the initial saturated state of condensing vapor in two ways
(Fig.~\ref{fig:twoLCL_scheme}). In the \emph{two-LCL estimate}, vapor
supplied by surface evaporation, with fraction $-Q_{e,s}/Q_c$, is
assigned the lifting condensation level (LCL) calculated from local
surface-air temperature, pressure, and humidity at each condensation
footprint. Vapor supplied by atmospheric evaporation, with fraction
$-Q_{e,a}/Q_c$, is assigned the LCL calculated from the local air
conditions at the atmospheric evaporation levels. The LCL is the
height where an adiabatically lifted parcel first becomes saturated.
Denoting the geometric-mean LCL temperatures weighted by positive
and negative SLH heating magnitudes by $T_{1,s}$ and $T_{1,a}$,
respectively, we estimate
\begin{equation}\label{eq:gpm-logT}
\begin{aligned}
\left\langle\ln\frac{T_1}{T}\right\rangle_c
\simeq{}&-\frac{Q_{e,s}}{Q_c}\ln\frac{T_{1,s}}{T_{\rm ref}}
-\frac{Q_{e,a}}{Q_c}\ln\frac{T_{1,a}}{T_{\rm ref}}\\
&-\left\langle\ln\frac{T}{T_{\rm ref}}\right\rangle_c,
\end{aligned}
\end{equation}
where $T$ is MERRA-2 temperature at each positive SLH level,
the last term is weighted by positive SLH heating, and the arbitrary
$T_{\rm ref}$ cancels. Equations~\eqref{Wv} and \eqref{Wg} then give
$W_v$ and $W_g$, with the latter using
\begin{equation}\label{eq:gpm-z1}
\langle z_1\rangle_c\simeq
-\frac{Q_{e,s}}{Q_c}z_{1,s}
-\frac{Q_{e,a}}{Q_c}z_{1,a}.
\end{equation}
Here $z_{1,s}$ and $z_{1,a}$ are the corresponding heating-weighted
mean LCL heights. 

For $z_{1,s}\simeq0.7$~km and
$z_{1,a}\simeq2.9$~km, we find
$\langle z_1\rangle_c\simeq1.2$~km
and $W_g\simeq1.5$~W~m$^{-2}$.

In the \emph{one-onset estimate}, we instead estimate
the mean initial saturation state of vapor supplied by
both surface and atmospheric evaporation using the
lowest level of positive SLH heating.
 In each footprint containing positive SLH heating,
we identify the lowest level at which $\dot T>0$.
The positive-heating-weighted mean of these footprintwise levels
is $z_{c,\min}$; $T_{c,\min}$ is the corresponding
positive-heating-weighted geometric mean of MERRA-2 temperatures.
We estimate
\begin{equation}\label{eq:gpm-onset}
\left\langle\ln\frac{T_1}{T_{\rm ref}}\right\rangle_c
\simeq\ln\frac{T_{c,\min}}{T_{\rm ref}},
\qquad
\langle z_1\rangle_c\simeq z_{c,\min}.
\end{equation}
The twelve-period means are $z_{c,\min}\simeq1.9$~km
and $T_{c,\min}\simeq284.6$~K.

We estimate $W_K$ directly from the instantaneous three-hourly
MERRA-2 pressure-level wind and geopotential height fields.
The local kinetic-energy generation per unit mass is
$c_K=-\mathbf u\cdot\nabla_p\Phi$, where $\Phi=gZ$ and
$Z$ is geopotential height (the MERRA-2 $H$ field).  We evaluate
horizontal geopotential gradients on pressure surfaces and integrate
$c_K$ over the atmospheric mass column.
For the unresolved surface layer, we interpolate $c_K$
linearly in pressure from zero at the surface to its
value at the lowest valid pressure level; alternative
extrapolations are detailed in Appendix~\ref{app:WK}.

OpenAI ChatGPT was used as an interactive assistant for manuscript editing, checking mathematical derivations, 
developing and verifying numerical analysis code, preparing figures, and reference verification.

\section{Results}
\label{sec:results}

\subsection{Kinetic energy generation $W_K$ from MERRA-2}
\label{sec:resultsWK}

\begin{figure*}[tbp]
\centering
\includegraphics[width=\textwidth]{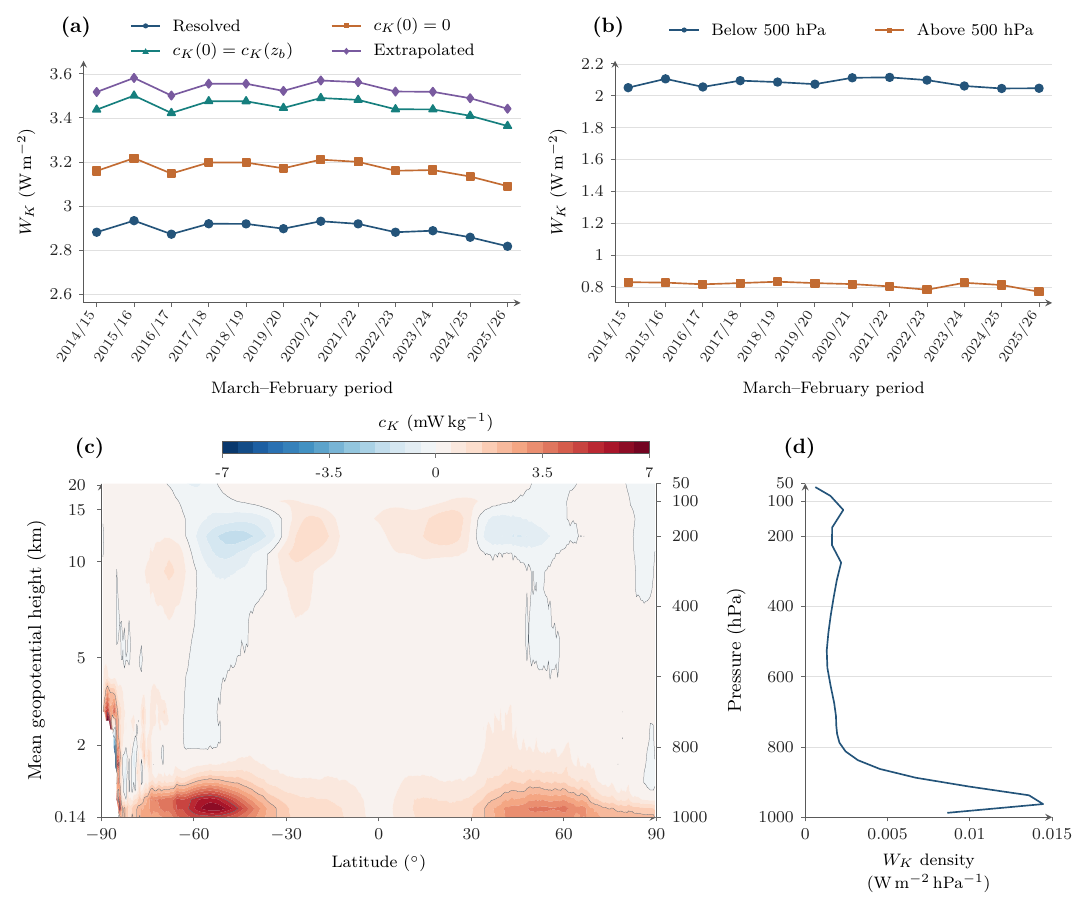}
\caption{
Kinetic energy generation diagnosed from MERRA-2. (a) Global $W_K$ for resolved pressure levels and three treatments of the unresolved surface layer. (b) Resolved $W_K$ below and above 500 hPa. (c) Latitude–pressure distribution of local kinetic energy generation $c_K$ (mW kg$^{-1}$), with corresponding mean geopotential heights. (d) Vertical profile of resolved $W_K$ density (W m$^{-2}$ hPa$^{-1}$). Panels (c,d) show the study-period means for 8 March 2014–28 February 2026, displayed over 50–1000 hPa.
}
\label{fig:WK}
\end{figure*}

The interannual variation in $W_K$ is relatively small
(Fig.~\ref{fig:WK}a). Over the twelve March--February periods,
the mean resolved power is $W_K=2.89\pm0.03$~W~m$^{-2}$
(mean $\pm$ interannual standard deviation), of which
$2.08\pm0.03$~W~m$^{-2}$ is generated below 500~hPa and
$0.81\pm0.02$~W~m$^{-2}$ above (Fig.~\ref{fig:WK}b).

Our resolved estimate exceeds the $2.46$~W~m$^{-2}$
reported by \citet{huang15} for the earlier MERRA
reanalysis over 1979--2010. The latter study used data
at a lower horizontal resolution
($1.25^\circ\times1.25^\circ$, compared with
$0.5^\circ\times0.625^\circ$ for MERRA-2).
More recently, \citet{oh2026lec} reported a value of
$2.87$~W~m$^{-2}$ from the Lorenz conversion terms
for MERRA-2 over 1980--2023, effectively identical
to our estimate. Their corresponding values for ERA5
and JRA-3Q are $2.45$ and $2.46$~W~m$^{-2}$,
respectively, approximately $0.4$~W~m$^{-2}$
(15\%) lower than for MERRA-2.

Our separate comparison for 2025 indicates that much
of this discrepancy arises above 500~hPa, where
MERRA-2 generates more kinetic energy than ERA5
and JRA-3Q
\citep[][data not shown here]{makarieva2026alphaomega}.
At these heights, positive and negative local $c_K$
partially cancel (Fig.~\ref{fig:WK}c), increasing the
uncertainty of the net estimate. Below 500~hPa,
kinetic energy generation is predominantly positive
(Fig.~\ref{fig:WK}c), with much less cancellation.
Furthermore, the reanalysis is more strongly
constrained by observations at these levels.
The lower-atmospheric contribution to $W_K$
is therefore more robustly estimated.

The vertical profile of $W_K$ density has a pronounced
maximum in the lower atmosphere (Fig.~\ref{fig:WK}d).
Thus, restricting the calculation to resolved pressure
levels omits the contribution of the surface layer and
underestimates the actual $W_K$.
We considered three treatments of this unresolved layer:
assuming $c_K=0$ at the surface yields
$W_K=3.17\pm0.04$~W~m$^{-2}$; holding $c_K$ constant
at its lowest resolved value gives
$3.45\pm0.04$~W~m$^{-2}$; and linear extrapolation
from the two lowest resolved levels gives
$3.53\pm0.04$~W~m$^{-2}$.
In subsequent analyses, we adopt the most conservative
of these estimates, $W_K=3.17$~W~m$^{-2}$,
corresponding to $c_K=0$ at the surface.
Overall, these results support a global kinetic energy
generation rate of approximately $3$~W~m$^{-2}$.

\subsection{Gravitational power of precipitation $W_P$ from GPM-SLH}
\label{sec:resultsWP}

\begin{figure}[htbp]
\centering
\includegraphics[width=0.97\linewidth]{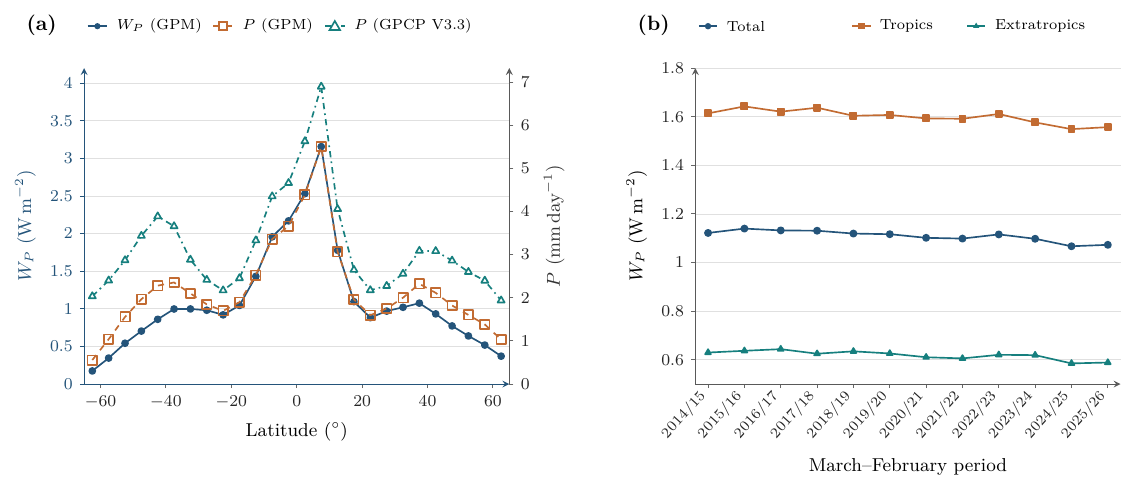}

\caption{
Gravitational power of precipitation $W_P$ estimated
from GPM/DPR spectral latent-heating profiles.
(a) Latitude profiles of $W_P$ and surface precipitation
$P$ from GPM, compared with $P$ from GPCP V3.3,
for March 2014--December 2025. All three quantities
are averaged over $5^\circ$ latitude bands.
(b) $W_P$ for twelve March--February periods
(2014/15--2025/26), normalized by the area of
the entire Earth (Total), the tropics
($30^\circ$S--$30^\circ$N), and the extratropics
($30^\circ$--$90^\circ$ in both hemispheres),
respectively. GPM observations cover
$65^\circ$S--$65^\circ$N; contributions poleward
of $65^\circ$ are not included, although the
corresponding areas are included in the
normalization.
}
\label{fig:WP}
\end{figure}

Our GPM-SLH analysis covers $65^\circ$S--$65^\circ$N,
which contains approximately 97\% of global precipitation,
according to our independent analysis of the GPCP
V3.3 monthly product \citep{huffman2025gpcp33}.
We normalize $W_P$ by the area of the entire Earth,
the tropics ($30^\circ$S--$30^\circ$N), and the
extratropics ($30^\circ$--$90^\circ$ in both hemispheres),
respectively. Contributions poleward of $65^\circ$
are omitted. Their contribution to global $W_P$
is expected to be small, since $W_P/P$ generally
decreases toward the poles (Fig.~\ref{fig:WP}a).

The mean gravitational power of precipitation,
normalized by the entire Earth's surface area, is
$W_P=1.11\pm0.02$~W~m$^{-2}$
(mean $\pm$ interannual standard deviation).
The corresponding regional means are
$1.60\pm0.03$~W~m$^{-2}$ in the tropics and
$0.62\pm0.02$~W~m$^{-2}$ in the extratropics
(Fig.~\ref{fig:WP}b).
The tropical estimate is close to the corrected
value of $1.5$~W~m$^{-2}$ reported by
\citet{pa12,pa12corr} using observations from 1999--2007.

For March 2014--December 2025, the mean GPM-derived
surface precipitation within $65^\circ$S--$65^\circ$N
is $2.36$~mm~day$^{-1}$, compared with
$3.31$~mm~day$^{-1}$ from GPCP V3.3
(Fig.~\ref{fig:WP}a), a difference of nearly 30\%.
Our GPM precipitation estimates come from the
Ku-band radar-based V07C SLH product, rather than
the combined radar--radiometer retrieval.
This distinction is important when considering
changes between product versions.
\citet{behrangi2024gpm} found that oceanic precipitation
estimated from GPM KuPR decreased by about 8\%
from V06 to V07, whereas the combined
radar--radiometer estimate increased by about 9\%.
Meanwhile, GPCP V3.3 reports 8.5\% higher global
precipitation than V3.2 over 1992--2023,
mainly due to increases over the oceans
\citep{behrangi2026gpcp}. Part of this increase
reflects the transition to the V07 combined
radar--radiometer retrieval.

However, the approximately 30\% discrepancy
in surface precipitation does not necessarily
imply an equally large underestimate of $W_P$.
As shown by \citet{jas13}, an underestimate
of surface precipitation confined to the lowest
kilometre contributes relatively little to
gravitational power: a 25\% underestimate of
precipitation in this layer corresponds to
only about a 5\% underestimate of $W_P$.
Thus, although the GPM--GPCP discrepancy introduces
uncertainty into our estimate of $W_P$, its
magnitude cannot be inferred directly from
the difference in surface precipitation.

Another source of uncertainty in our $W_P$
estimate is the treatment of SLH profiles without
separately accounting for melting and freezing.
Our analysis of the microphysical heating profiles
in Fig.~2-1 of \citet{tao2016lh} indicates that
excluding these processes reduces $W_P$ by about
2\% in convective clouds but increases it by about
12\% in stratiform clouds (data not shown).
The latter effect may contribute to an underestimate
of $W_P$, particularly at higher latitudes.
Overall, an uncertainty in $W_P$ of order
20\% appears plausible.

Taking $W_P\simeq1.1$~W~m$^{-2}$ as a
representative global estimate, together with
$W_K\simeq3.2$~W~m$^{-2}$, we obtain
a total atmospheric power under the hydrostatic
approximation of approximately
$W=W_K+W_P\simeq4.3$~W~m$^{-2}$,
with a likely uncertainty of order 20\%.

\subsection{Steam-engine power $W_v=W_g+W_c$ from GPM-SLH and MERRA-2}
\label{sec:resultsWv}

\begin{figure}[htbp]
\centering
\includegraphics[width=0.97\linewidth]{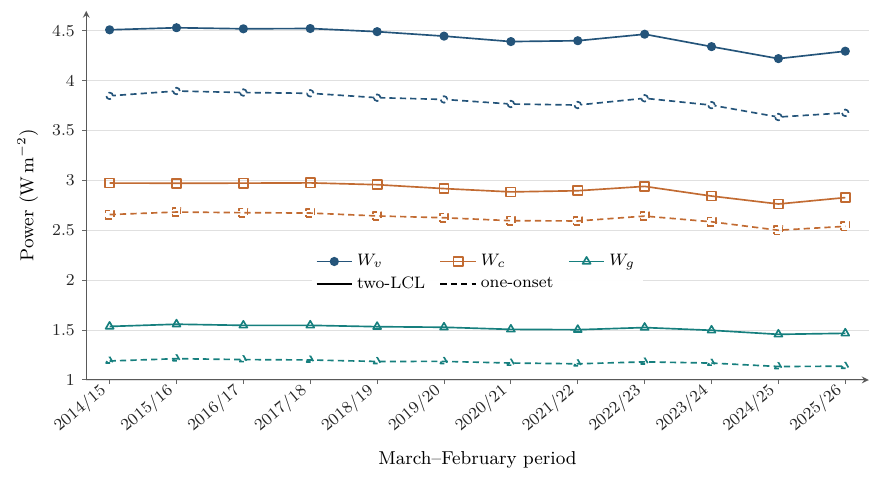}

\caption{
Steam-engine expansion power $W_v$ and its
gravitational ($W_g$) and condensation-related
($W_c$) components estimated from GPM/DPR
spectral latent heating and MERRA-2 using the
two-LCL (solid lines) and one-onset (dashed lines)
methods. The values are normalized by the area
of the entire Earth for twelve March--February
periods (2014/15--2025/26). The GPM analysis
covers $65^\circ$S--$65^\circ$N; contributions
poleward of $65^\circ$ are omitted, while the
polar areas are included in the normalization.
}
\label{fig:Wv}
\end{figure}

Besides precipitation uncertainties discussed in
Section~\ref{sec:resultsWP}, a major source of uncertainty in
$W_v$ and its components is the estimation of $z_1$ and $T_1$,
the height and temperature at which vapor that subsequently
condenses first enters saturation.
The two-LCL method assumes
undiluted ascent to saturation from the corresponding evaporation
levels (Fig.~\ref{fig:twoLCL_scheme}).
This assumption is consistent with the model-derived
microphysical profiles of \citet[][their Fig.~2-1]{tao2016lh},
which show gross condensation extending down to
approximately 0.3~km above the surface in both convective
and stratiform regions, even though evaporation dominates
over much of the lower stratiform atmosphere.
Accounting for atmospheric evaporation, the two-LCL method
yields a condensation-weighted mean initial saturation height
$\langle z_1\rangle_c\simeq1.18$~km.

Over the twelve March--February periods, the two-LCL
method gives $W_v=4.43\pm0.10$~W~m$^{-2}$,
comprising $W_g=1.52\pm0.03$~W~m$^{-2}$ and
$W_c=2.91\pm0.07$~W~m$^{-2}$
(Fig.~\ref{fig:Wv}; mean $\pm$ interannual
standard deviation).

An alternative, more conservative estimate is based on
the lowest level of positive retrieved net latent heating
in each footprint, whose positive-heating-weighted mean
is $z_{c,\min}$.
The corresponding
heights and MERRA-2 temperatures are used to estimate the
mean effective initial saturation state for vapor originating
from both surface and atmospheric evaporation. 
Because net evaporation can dominate the lower cloud layers
and mask gross condensation there, we expect the lowest
level of positive net latent heating to lie above the
onset of gross condensation.

The one-onset method gives
$z_{c,\min}\simeq1.9$~km and an effective
initial saturation temperature
$T_{c,\min}\simeq284.6$~K, compared with
the effective geometric-mean temperature
$T_1\simeq287.1$~K for the two-LCL method.

The resulting powers are
$W_v=3.80\pm0.08$~W~m$^{-2}$,
$W_g=1.18\pm0.02$~W~m$^{-2}$ and
$W_c=2.62\pm0.06$~W~m$^{-2}$
(Fig.~\ref{fig:Wv}). These estimates are lower than the
corresponding two-LCL values by approximately 14\%, 22\%
and 10\%, respectively.

The arithmetic means of the two estimates are
$W_v\simeq4.11$~W~m$^{-2}$,
$W_g\simeq1.35$~W~m$^{-2}$ and
$W_c\simeq2.76$~W~m$^{-2}$.

Using the Earth-area-normalized condensation
heating $Q_c\simeq74$~W~m$^{-2}$, or
$P_c=Q_c/L\simeq2.96\times10^{-5}$~
kg~m$^{-2}$~s$^{-1}$,
and $T_c\simeq271$~K, we obtain from
Eq.~\eqref{kappa_eff}
$\kappa_{\rm eff}\simeq-0.21$ for the two-LCL method
and $\kappa_{\rm eff}\simeq-0.29$ for the one-onset method.
These correspond to effective incompleteness-of-condensation
ratios $\gamma_2/\gamma_1\simeq0.08$ and $0.12$,
respectively, or effective degrees of condensation
$1-\gamma_2/\gamma_1\simeq0.92$ and $0.88$.

\section{Discussion}
\label{sec:discussion}

\subsection{Numerical results}

Under the hydrostatic approximation, we estimated
atmospheric mechanical power as $W=W_K+W_P$,
diagnosing $W_K$ from MERRA-2 and $W_P$
separately using GPM-SLH and MERRA-2 atmospheric fields.

The gravitational power of precipitation $W_P$ was
already identified and estimated by \citet[p.~30]{gorshkov1995},
who listed it separately from wind power among terrestrial
energy flows. The distinction between mechanical work
required to lift precipitation and kinetic energy generation
by atmospheric motions was subsequently discussed by
\citet{pauluis2000} and \citet{pa12}, the latter providing
the first satellite-based estimate of precipitation-induced
dissipation using TRMM observations. However, to our
knowledge, the general decomposition $W=W_K+W_P$
from the atmospheric continuity equation has not been
derived in the peer-reviewed literature.
The conventional equivalence
between total pressure work $W$ calculated by integrating
$-(1/\rho)Dp/Dt$ over atmospheric mass and horizontal
pressure-gradient work $W_K$ calculated by integrating
$-\mathbf{u}\cdot\nabla_h p$ over atmospheric volume,
which neglects the contribution $W_P$ arising from
gaseous-mass sources and sinks, continues to be assumed
in recent studies \citep{wangyang2023lec,oh2026lec}.\footnote{An
earlier derivation of $W=W_K+W_P$ \citep{makarieva2017powerpreprint}
remained unpublished in a peer-reviewed journal. Our recent preprint
\citep{makarieva2026alphaomega} reviews the continued identification
of $W$ with $W_K$ in studies of the Lorenz energy cycle.}

Our conservative surface-layer treatment gives
$W_K=3.2$~W~m$^{-2}$ ($2.9$~W~m$^{-2}$ in the resolved
layers; Section~\ref{sec:resultsWK}).
Differences of about 15\% among reanalyses, largely
concentrated in the upper atmosphere \citep{oh2026lec},
together with the uncertainty of the unresolved surface
layer, suggest a representative $W_K\simeq3$~W~m$^{-2}$
with an uncertainty of order 20\%.

We obtain $W_P=1.1$~W~m$^{-2}$
(Section~\ref{sec:resultsWP}), with a plausible uncertainty
of order 20\% arising primarily from satellite retrievals
of precipitation and latent heating, possible underestimation
of near-surface precipitation, and the treatment of melting
and freezing. Together, the reference estimates give
$W=W_K+W_P\simeq4.3$~W~m$^{-2}$, also with an uncertainty
of order 20\%. The smaller interannual standard deviations
reported in the Results do not represent these observational
and methodological uncertainties.

Kinetic energy generation is concentrated in the lower
atmosphere (Fig.~\ref{fig:WK}b-d). We denote the
contributions below and above 500~hPa by $W_{K,l}$
and $W_{K,u}$, respectively. Below 500~hPa,
$W_{K,l}=2.1$~W~m$^{-2}$ in the resolved layers,
increasing to $2.4$~W~m$^{-2}$ with the conservative
surface-layer correction. The upper atmosphere contributes
$W_{K,u}=0.8$~W~m$^{-2}$, or approximately one quarter
of the corrected total.

The relatively small positive net contribution aloft
results from partial cancellation between positive and
negative kinetic energy generation, including negative
work where air moves against temperature-related
pressure gradients, as in the Ferrel cells
(Fig.~\ref{fig:WK}c). In contrast, lower-atmospheric
kinetic energy generation is predominantly positive,
with a pronounced near-surface maximum
(Fig.~\ref{fig:WK}c,d).

Our separately diagnosed positive saturated water-vapor
expansion power is $W_v=4.4$~W~m$^{-2}$ with the
two-LCL method and $3.8$~W~m$^{-2}$ with the
one-onset method (Section~\ref{sec:resultsWv}).
The arithmetic mean is $W_v\simeq4.1$~W~m$^{-2}$,
with a representative uncertainty of order 30\%.

Water-vapor expansion power consists of two physically
distinct contributions. The gravitational component $W_g$
arises because condensation removes gaseous mass during
ascent, so that more gas molecules ascend and expand
than descend and contract. This component is quantified
by the hydrostatic decrease in total-air pressure during
ascent (Eq.~\eqref{Wg}).

The gravitational power of precipitation $W_P$ likewise
arises from this imbalance, but differs from $W_g$ in
two respects. First, $W_g$ considers expansion only
between $z_1$ and $z_c$, with $z_1>z_e$, whereas
$W_P$ considers the full interval from evaporation
at $z_e$ to condensation at $z_c$
(Fig.~\ref{fig:twoLCL_scheme}). Second, $W_g$ is
proportional to the molecular mass of air through
the hydrostatic pressure decrease. 
Thus, if $z_1$ and $z_e$ coincided, we would obtain
$W_g\simeq(M/M_v)W_P=1.6W_P$, rather than
the present $W_g\simeq1.2W_P$, based on the mean
of the two estimates of $W_g$.
The excess ascending
water-vapor molecules are lighter than the surrounding
air, contributing to buoyancy and hence to greater
positive work.

The condensation-related component $W_c$ arises from
the additional decrease in saturated water-vapor partial
pressure relative to total-air pressure during condensation
(Eqs.~\eqref{dg} and \eqref{Wc_gamma}).
For the two-LCL and one-onset methods, $W_c=2.9$
and $2.6$~W~m$^{-2}$, respectively, with a mean
$W_c\simeq2.8$~W~m$^{-2}$.
Both estimates are comparable in magnitude to, and exceed,
the estimated lower-atmospheric kinetic energy generation,
$W_{K,l}\simeq2.4$~W~m$^{-2}$.

\begin{figure}[htbp]
\centering
\includegraphics[width=0.97\linewidth]{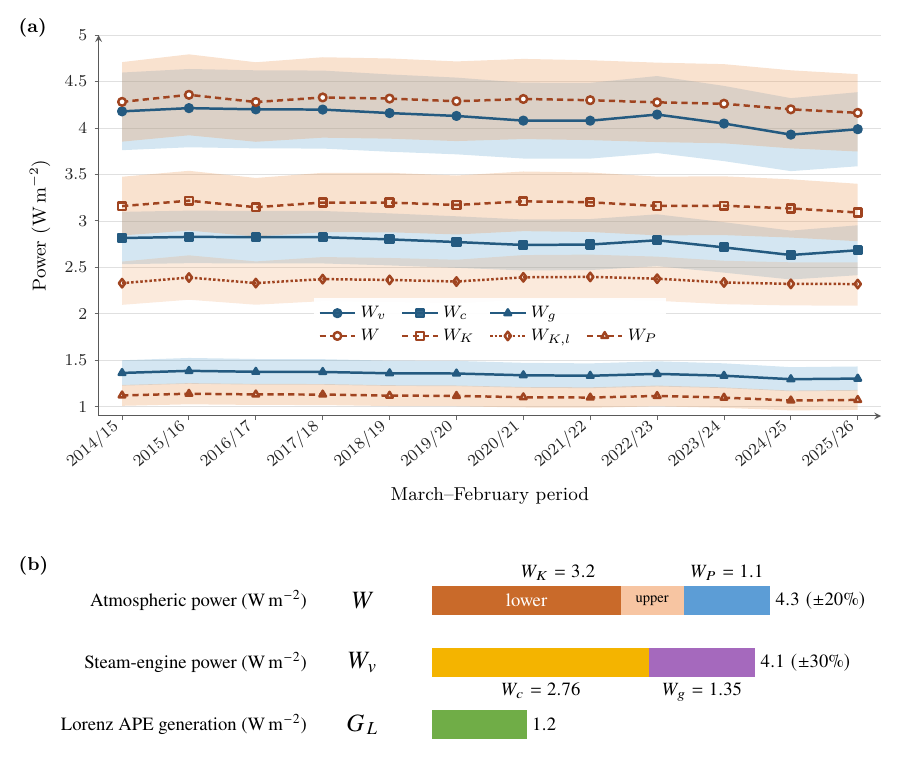}

\caption{%
(a) Comparison of steam-engine expansion power $W_v$ and its
condensation-related ($W_c$) and gravitational ($W_g$) components
with atmospheric mechanical power $W=W_K+W_P$, kinetic energy
generation $W_K$ and its lower-atmospheric contribution $W_{K,l}$
(below 500~hPa, including the surface-layer correction), and gravitational
power of precipitation $W_P$. Blue solid lines show arithmetic means
of the two-LCL and one-onset estimates; orange broken lines show
separately diagnosed atmospheric powers. Shaded areas indicate
$\pm10\%$ of each series for visual comparison.
(b) Corresponding period-mean power budgets, with $W_K$ divided into
lower- and upper-atmospheric contributions.
The labels $\pm20\%$ for $W$ and $\pm30\%$ for $W_v$ indicate
representative uncertainties.
For comparison, the green bar shows the global Lorenz
available potential energy (APE) generation
$G_L\simeq1.2~\mathrm{W\,m^{-2}}$
estimated by \citet{romanski2013} from predominantly
observation-based diabatic heating and temperature
fields for 1997--2000.}
\label{fig:WvW}
\end{figure}

Figure~\ref{fig:WvW} shows that $W$ and $W_v$
differ by approximately 4\%, while $W_K$ and
$W_c$ differ by approximately 13\%.
Importantly, $W_c\simeq2.8$~W~m$^{-2}$ is also
close to the lower-atmospheric
$W_{K,l}\simeq2.4$~W~m$^{-2}$,
whose physical relationship to $W_c$ we consider
below. These differences are smaller than the
estimated uncertainties of 20--30\%.

Our steam-engine analysis concerns the positive expansion
work of saturated water vapor, which is energetically
linked to latent heat release during condensation
through the Clausius--Clapeyron relation.
In a related but different approach,
\citet{pauluis2002b} evaluated the global integral of
$p_v\nabla\cdot\mathbf v$, likewise distinguishing
the hydrostatic decrease of total-air pressure from
changes in water-vapor abundance due to phase transitions.
However, because their integral also included evaporation,
the positive condensation-related contribution proportional
to $RT_c$ was more than offset by the negative
evaporation-related contribution proportional to $RT_e$,
with $T_e>T_c$ ($T_e/T_c\simeq1.08$ in our data;
Fig.~\ref{fig:twoLCL_scheme}).
Thus, the condensation and evaporation contributions
largely cancelled in their global integral, which
remained dominated by $(M/M_v)W_P$.
This cancellation does not imply that $W_c$ vanishes:
their global integral does not separately evaluate
the condensation-related expansion work defined here,
and its potential contribution to atmospheric power
generation was not investigated in that framework.

\subsection{Relation to condensation-induced atmospheric dynamics}
\label{sec:ciad}

In contrast, the condensation-related contribution to
atmospheric power has been central to
condensation-induced atmospheric dynamics (CIAD)
from the outset, although its relationship to the
expansion work of the water-vapor steam engine had
not previously been established.

CIAD originated from the idea that the vertical
distribution of water-vapor partial pressure maintained
by condensation differs from that associated with the
hydrostatic decrease of total-air pressure, and that
this difference can generate atmospheric power
proportional to the condensation rate
\citep[for a review, see][]{makarieva19}.

The proposed local power density is
\begin{equation}\label{Wloc}
\begin{aligned}
\sigma_{\rm CIAD}
&=-w\left(\frac{\partial p_v}{\partial z}
 -\frac{p_v}{p}\frac{\partial p}{\partial z}\right)\\
&=-\frac{wp}{(1+\gamma)^2}\frac{\partial\gamma}{\partial z}
\simeq -wp\frac{\partial\gamma}{\partial z}.
\end{aligned}
\end{equation}

Using water-vapor and dry-air continuity, neglecting
vapor diffusion, and assuming that vertical advection
dominates the material change of $\gamma$, we obtain
$\sigma_{\rm CIAD}\simeq RTC/M_v$ to leading order in
the vapor fraction, where $C$ is the local condensation
rate in kg~m$^{-3}$~s$^{-1}$. Integration over the
condensation region $\Omega_c$ gives
\begin{equation}\label{WcCIAD}
\begin{aligned}
W_{\rm CIAD}
&\equiv\frac{1}{\mathcal S}
\int_{\Omega_c}\overline{\sigma_{\rm CIAD}}\,d\Omega\\
&\simeq\frac{RP_cT_c}{M_v}
=\frac{W_c}{1+\kappa_{\rm eff}},
\end{aligned}
\end{equation}
where the final equality follows from
Eq.~\eqref{kappa_eff}. Thus $W_{\rm CIAD}$ corresponds
to $W_c$ in the complete-condensation limit
$\kappa_{\rm eff}\to0$. For the two-LCL and one-onset
estimates, $\kappa_{\rm eff}\simeq-0.21$ and $-0.29$,
respectively, so that $W_c$ is reduced relative to the
integrated CIAD expression by factors of approximately
$0.79$ and $0.71$.

If $\sigma_{\rm CIAD}$ is identified, in order of magnitude,
with local kinetic energy generation by horizontal pressure
gradients, $\sigma_{\rm CIAD}\sim-\mathbf u\cdot\nabla_h p$,
dimensional analysis gives a characteristic horizontal
pressure difference. Taking $u/w\sim L/H$,
$|\nabla_h p|\sim\Delta p/L$, and
$|\partial\gamma/\partial z|\sim\Delta\gamma/H$,
we obtain $\Delta p\sim p\Delta\gamma$.

For nearly complete condensation,
$\Delta\gamma\sim\gamma_1$, giving
$\Delta p\sim p\gamma_1$, of the order of the initial
saturated water-vapor partial pressure, typically
about 10~hPa in the lower atmosphere. This is
comparable to pressure differences in large-scale
circulations ($L\sim10^3$~km) \citep{hess07},
including about 12~hPa across the Hadley circulation
\citep{tellus17}. \citet[p.~4]{holton04} also noted pressure differences
of order 10~hPa in systems with very different spatial
scales, including tornadoes, squall lines, and hurricanes.

Within CIAD, this relative constancy is interpreted as a
consequence of the constraint that condensation imposes
on kinetic energy generation. For tropical cyclones,
equating the condensation-domain integral of
$\sigma_{\rm CIAD}$ to boundary-layer kinetic energy
generation and applying cyclostrophic balance yields an
upper limit on maximum wind velocity \citep{mn25}.

\subsection{Why does atmospheric power approximately coincide with steam-engine power?}
\label{sec:discussion-why}

The comparison posed in the Introduction can be
written as
\begin{equation}
W_v-W=(W_g-W_P)+(W_c-W_K).
\label{eq:why-power-difference}
\end{equation}
With our global estimates, $W_c\simeq2.8$~W~m$^{-2}$
is close to both the total $W_K\simeq3.2$~W~m$^{-2}$
and the lower-atmospheric $W_{K,l}\simeq2.4$~W~m$^{-2}$.
The mechanism we propose primarily concerns the
latter correspondence. We seek to explain how
condensation-related power generates kinetic energy
in the lower atmosphere.
We leave the smaller, consistently positive difference
$W_g-W_P\simeq0.2$~W~m$^{-2}$ and its possible
connection to the observed upper-atmospheric
kinetic energy generation
$W_{K,u}\simeq0.8$~W~m$^{-2}$
for future research.

We propose that the condensation-related steam-engine power $W_c$ represents the rate of transformation of potential energy associated with the non-equilibrium vertical distribution of saturated water vapor in the Earth's gravitational field into potential energy associated with surface pressure gradients. As a mechanical energy transformation rather than a conversion of heat into work, it involves no additional Carnot-efficiency factor. The resulting potential energy is converted into kinetic energy by cross-isobaric air flow in the frictional lower atmosphere.

Condensation thus offers the entire air mixture an opportunity
to expand toward the lower surface pressure it maintains.
However, while condensed water does not require compression,
the expanded air does.
In a horizontally isothermal atmosphere, upper-level compression
would nearly cancel the positive pressure work of the
lower-level expansion. This compensation is illustrated
using a simplified mass-related Lorenz-type available-energy
formulation in Appendix~\ref{app:Lorenz}.

Here expansion and compression refer to horizontal
pressure work. Under hydrostatic equilibrium,
vertical pressure-gradient forces are balanced by gravity,
with the gravitational contributions to steam-engine
and atmospheric power accounted for separately by
$W_g$ and $W_P$, respectively.

\begin{figure}[htbp]
\centering
\includegraphics[width=0.93\linewidth]{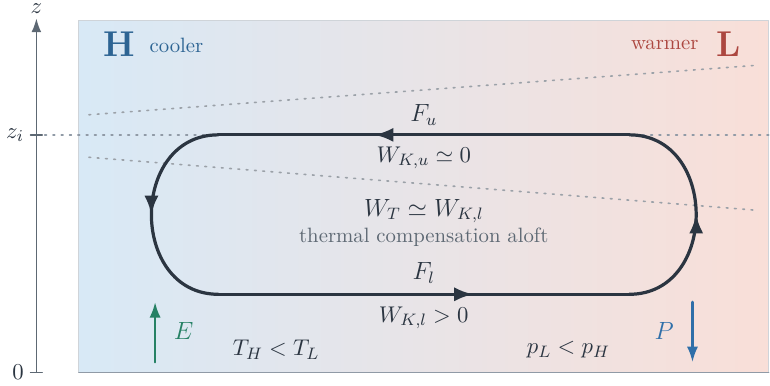}
\caption{Idealized circulation with lower-level flow
from the cooler high-pressure region $H$ to the warmer
low-pressure region $L$ and a return flow aloft.
Dotted lines indicate isobars; $z_i$ is the isobaric
height. $E$ and $P$ denote net evaporation and
precipitation. For $W_{K,u}\simeq 0$, thermal
compensation implies $W_T\simeq W_{K,l}$.}
\label{fig:airflow-isobaric}
\end{figure}

The crucial difference between the lower and upper branches (Fig.~\ref{fig:airflow-isobaric})
is that the frictional surface flow requires appreciable
pressure work, whereas the return flow aloft can proceed
with negligible pressure work. We propose that the
condensation-driven surface flow unlocks differential heating
that warms the upper-level return flow, modifying the
pressure field and preventing compression\footnote{
A lower molar mass of the ascending air can likewise
contribute to offsetting the compression.}.
The return flow thus becomes nearly isobaric,
with $W_{K,u}\simeq0$.

Here, ``unlocks'' means that we propose that the same
horizontal temperature differences would be unable to drive a
comparable circulation without the condensation-related
power input at the surface.

Schematically, in the idealized case $W_{K,u}\simeq0$,
the negative surface-pressure contribution to upper-level
work, approximately $-W_{K,l}$ in the isothermal
limit, is offset by the positive temperature-related
contribution $W_T$. Thus, as illustrated in
Fig.~\ref{fig:airflow-isobaric},
$W_K\simeq W_{K,l}\simeq W_T$:
the same net work can be described through its
lower-level pressure-work origin or through the
differential heating that prevents its cancellation aloft
(Appendix~\ref{app:Lorenz},
Eq.~\eqref{eq:Wc-WT-Lorenz}).
In the idealized case considered here, when
$W_{K,l}\simeq W_c$, the condensation-related
steam-engine power and temperature-related power
coincide: $W_T\simeq W_c$.

Notably, $W_T$ depends not on the horizontal temperature
difference $\Delta T$ alone, but on its product with the
mean height of the upper flow. Thus, for a given surface
pressure gradient, the return flow can be accommodated
by a range of horizontal temperature gradients, depending on its
height (Eq.~\eqref{eq:zDeltaT}).
This flexibility supports the proposed ``unlocking''
mechanism.

The foregoing argument considered an idealized case
with $W_{K,u}\simeq0$ and $W_{K,l}\simeq W_c$.
In our global estimates, by contrast,
$W_{K,u}\simeq0.29W_c$, while
$W_{K,l}\simeq0.86W_c$.

The difference between $W_{K,l}$ and $W_c$
may reflect the less-than-unity efficiency of the
mechanical transformation of steam-engine expansion
work into the potential energy associated with surface
pressure gradients. One factor affecting this efficiency
may be the spatial and temporal overlap between
evaporation and condensation. Where these processes
overlap in space and time, their opposing contributions
to the gaseous-mass budget partially cancel, reducing
the column-integrated net mass source or sink.
As discussed in Appendix~\ref{app:Lorenz},
mass-related available-energy generation depends
on the covariance between this column-integrated
net mass source and surface pressure, rather than
on the condensation rate alone.
Such overlap may therefore reduce the power
available for kinetic energy generation in the
lower atmosphere. This possibility requires
further study.

\subsection{Efficiency and mechanical power}
\label{sec:efficiency}

Previous studies have obtained substantially reduced,
but quantitatively different, estimates of the mechanical
work associated with atmospheric moisture.
\citet{pauluis2002b}, evaluating the global integral of
$p_v\nabla\cdot\mathbf{v}$, estimated water-vapor work
to be approximately 20--30\% of the work of a perfect
heat engine transporting the same latent heat.
Using a different thermodynamic formulation,
\citet{pa11} found that the efficiency of a fully
unsaturated steam cycle under Earth-like conditions
is approximately one-sixth to one-fifth of Carnot
efficiency, increasing to Carnot efficiency for a
fully saturated cycle.
These estimates concern different quantities and
assumptions, and their numerical values cannot be
directly applied to the relationship between $W_c$
and $W_{K,l}$ considered here.

We agree with the general conclusion that the net
mechanical work associated with moist atmospheric
cycles can be considerably smaller than the Carnot
benchmark. In our framework, two distinct reductions
can be identified. First, the positive water-vapor
expansion power corresponds to a fraction
$W_v/Q_c\simeq0.056$ of the latent heat released
during condensation, where
$Q_c\simeq74$~W~m$^{-2}$.
For the two-LCL and one-onset methods, this fraction
is approximately $0.060$ and $0.051$, respectively.
According to Eq.~\eqref{Wv},
\begin{equation}
\frac{W_v}{Q_c}\simeq
\left\langle\ln\frac{T_1}{T}\right\rangle_c,
\end{equation}
which approaches the corresponding Carnot-like
efficiency in the small temperature-difference limit.
Here $T_1$ is the initial saturation temperature,
rather than necessarily the evaporation temperature.
The commonly cited Carnot efficiency of the atmospheric
heat engine is approximately $0.1$, based on large-scale
atmospheric temperature contrasts
\citep{peixoto1992,pa12,bohren2023}.
This comparison reflects different temperature contrasts
and should not be interpreted as a measure of irreversible
efficiency loss.

Second, in the horizontally isothermal case, a further
and potentially much larger reduction occurs because
positive lower-level pressure work is almost entirely
offset by negative pressure work in the return flow
aloft. Consequently, $W_K\ll W_c$, and the net
mechanical efficiency relative to the latent-heat
input is much smaller than even the Carnot efficiency
associated with the initial-saturation--condensation
temperature difference. In the non-isothermal case,
however, under the condition of a small upper-level
pressure-work residual, $W_K\simeq W_c\simeq W_T$:
the net power can also be expressed in terms of
differential heating, including its sensible-heat
contribution, whose conversion efficiency can in
principle approach the Carnot limit.
The condensation-related and temperature-related
expressions characterize the same mechanical power
rather than additive contributions. Thus, a potentially
efficient contribution from differential heating can
coexist with a low net efficiency of the work
attributable to latent heat alone.

The dominant role of condensation that we propose lies
not in increasing thermodynamic efficiency, but in
making a circulation of substantial mechanical power
possible in the first place. Thus, a moist atmosphere
may have both lower thermodynamic efficiency and higher
mechanical power than a dry atmosphere.

Hertz suggested that in a dry atmosphere, temperature
differences ``would by themselves give rise merely to
movements of minor significance'' \citep{mulligan1997}.
In our interpretation, without the water cycle,
differential heating, which itself depends on air motion
and manifests itself in temperature-related pressure
gradients aloft, would be unable to generate comparable
surface pressure gradients and cross-isobaric flow
in the lower atmosphere.

\subsection{Disentangling causality}

The observed correspondence between $W_c$ and $W_{K,l}$,
together with the approximate equality $W_K\simeq W_T$,
does not by itself establish causality.
Differential heating might generate $W_K$
independently of the steam engine,
rather than the steam engine unlocking the
differential heating needed for the return flow.
In the former case, the proximity of $W_{K,l}$
and $W_c$ could be attributed to chance.
We can examine these alternatives in circulations
with contrasting upper-level pressure work.

Hurricanes offer such a contrasting test. They develop
over a largely isothermal surface, and their upper
outflow can generate negative pressure work, unlike
the small positive global-mean upper-atmospheric
contribution (Fig.~\ref{fig:WK}b,c).
Such motion against a significant pressure gradient
is possible because hurricanes convert a substantial
fraction of the pressure potential energy into kinetic
energy in the low-level inflow.
This fraction can reach about 40\%, compared with
about 2\% in large-scale, friction-dominated
circulations such as Hadley cells.\footnote{
The conversion fraction is
$k=\rho V^2/(2\Delta p_s)$, where $V$ is air velocity.
For hurricanes, $\Delta p_s$ is the surface pressure
drop from the outer storm environment to the radius
of maximum wind. Using the pressure profile of
\citet{holland1980}, gradient-wind balance in the
negligible-Coriolis limit gives
$k=b/[2(e-1)]\simeq0.36$ for a typical
Holland-profile parameter $b=1.25$.
For Hadley cells, $\Delta p_s\simeq10$~hPa,
$V\simeq7$~m~s$^{-1}$ and
$\rho\simeq1$~kg~m$^{-3}$ give $k\simeq0.025$.
}

In the modeled hurricane of \citet{smith18},
the positive and negative horizontal pressure-work
contributions at about day~3 were approximately
$+55$ and $-45$~TW, leaving $10$~TW net
within a cylinder of radius 300~km and height 20~km
(their Fig.~4; values estimated from volume-mean
terms). Thus, negative work in the upper outflow
consumes much of the kinetic energy generated
by positive pressure work in the low-level inflow.

We can compare the maximum local surface
dissipation in hurricanes with condensation-related
expansion power estimated from the maximum
eyewall precipitation rate, using the former as
a proxy for local kinetic energy generation.
According to \citet{emanuel1999}, an average
Atlantic hurricane with a maximum wind speed
$V_{\max}=50$~m~s$^{-1}$ dissipates kinetic
energy at the surface at a local rate
$D=\rho C_D V^3$, where
$\rho=1$~kg~m$^{-3}$ is air density and
$C_D=2\times10^{-3}$ is the drag coefficient.
The corresponding maximum local dissipation is
$D_{\max}\simeq250$~W~m$^{-2}$.

Independently, \citet{yang2021tc} obtained
a maximum azimuthally averaged eyewall precipitation
rate of $9.1$~mm~h$^{-1}$ for Category~3 hurricanes,
or about $220$~mm~day$^{-1}$.
For an illustrative condensation temperature
$T_c=270$~K, assuming complete condensation
($\kappa=0$), we obtain
\begin{equation}
\frac{W_c}{P_c}\simeq\frac{RT_c}{M_v}
\simeq1.25\times10^5\ \mathrm{J\,kg^{-1}},
\qquad
W_c\simeq320\ \mathrm{W\,m^{-2}}.
\label{eq:why-wc-hurricane}
\end{equation}

The two estimates are within 30\% of each other,
although they represent characteristic local maxima
rather than collocated measurements.
Nevertheless, similar power scales in circulations
with contrasting upper-level energetics provide
an additional consistency check for the proposed
condensation-driven lower-atmospheric pathway.
Even when, as in hurricanes, horizontal temperature
gradients associated with latent heat release
are insufficient to produce positive kinetic energy
generation aloft, the lower atmosphere still
generates power close to $W_c$.
More detailed considerations of hurricane energetics
within the CIAD framework are given by
\citet{pla11,mn25}.
	
\subsection{Implications of neglecting the second equality
in $W_K\simeq W_T\simeq W_c$}

The relatively small net kinetic energy generation
aloft (Fig.~\ref{fig:WK}d) creates a peculiar
situation: kinetic energy is generated predominantly
in the lower atmosphere by surface pressure gradients,
yet the same power can be expressed in terms of
horizontal temperature gradients,
$W_K\simeq W_T$.
Our results establish an additional correspondence:
$W_K\simeq W_c$, where $W_c$ is the
condensation-related power of the steam engine.

In view of this second correspondence, distinguishing
cause from effect is crucial for predicting
atmospheric responses to different forcings.
In a statistically steady atmosphere, kinetic energy
generation must balance frictional dissipation.
Turbulent by nature, this dissipation is unknown
a priori. We therefore cannot predict how
an Earth-like planet would circulate its air
from first principles alone.

Current parameterizations of turbulent friction
are based on observed relationships between wind
velocities, atmospheric state variables, and
surface properties. For the contemporary atmosphere,
models employing these parameterizations reproduce
the observed circulation, in which
$W_c\simeq W_T$ under the existing coupling between
differential heating and condensation.
Reproducing this circulation, however, does not
establish which of these processes is the primary
driver. Extrapolating model predictions to scenarios
in which this coupling changes requires correctly
identifying the primary determinant of circulation
and ensuring that its role is properly represented
in the model.
To give a practical example, a model that attributes
circulation primarily to differential heating may
predict that Amazon deforestation enhances atmospheric
moisture import by increasing the land--ocean temperature
contrast. In a global storm-resolving simulation
of complete Amazon deforestation, \citet{yoon2025}
found that circulation changes enhanced moisture
convergence, largely compensating for reduced
evapotranspiration, whereas most CMIP6 models
in a global-deforestation experiment predicted
reduced Amazon precipitation \citep{luo2022}.

An incomplete representation of $W_c$, combined
with an emphasis on $W_T$, may also contribute
to differences between modeled and observed
atmospheric circulations under present-day
conditions. For example, in a storm-resolving
configuration of ICON, \citet{paccini2023}
found excessive afternoon precipitation and
an underestimated and delayed secondary
early-morning rainfall maximum.
We suggest that examining the relative roles
of $W_c$ and $W_T$ could provide an additional
perspective on such discrepancies, particularly
the excessive concentration of rainfall during
hours of strong daytime heating.

These distinctions are particularly important
for predicting the effects of land-cover change.
If condensation determines whether circulation
takes place, reducing native forest cover
and hence the moisture supply could weaken
circulation despite increasing temperature contrasts.
Thus, different assumptions about the physical origin
of the power dissipated by turbulent friction may
lead to contrasting predictions under the same forcing.
More generally, the spatial reorganization of
evaporation and precipitation could have different
predicted effects on atmospheric circulation depending
on whether the mechanical role of condensation is
recognized.

The evidence presented here for the correspondence
between lower-atmospheric kinetic energy generation
and condensation-related expansion work of water vapor
suggests that this mechanism deserves explicit
consideration in understanding and predicting
atmospheric responses to environmental change.

\appendix
\section{Gravitational power of precipitation}
\label{app:WP}

Hydrostatic equilibrium, $\partial_z p=-\rho g$, with $\rho$ the
gaseous-air density, gives the time-mean vertical contribution to
pressure work:
\begin{equation}\label{WPapp1}
W_P\equiv-\frac{1}{\mathcal S}\int_\Omega
\overline{w\frac{\partial p}{\partial z}}\,d\Omega
=\frac{g}{\mathcal S}\int_\Omega\overline{\rho w}\,d\Omega.
\end{equation}

The nonnegative local net condensation and evaporation rates $C$ and
$E$ are defined in Section~\ref{sec:CCTc}. Surface evaporation is
included in $E$ as a source in a thin near-surface atmospheric layer.
The instantaneous gaseous-air continuity equation is
\begin{equation}\label{Ssource}
\frac{\partial\rho}{\partial t}
+\nabla\cdot(\rho\mathbf v)=E-C.
\end{equation}
Multiplying by height $z$ and using $\mathbf v\cdot\nabla z=w$
gives
\begin{equation}\label{height_continuity}
\frac{\partial(z\rho)}{\partial t}
+\nabla\cdot(z\rho\mathbf v)=\rho w-z(C-E).
\end{equation}
For a statistically steady atmosphere, the time-mean storage term
$\overline{\partial(z\rho)/\partial t}$ vanishes. Assuming also that
the gaseous-air mass flux through the boundary of $\Omega$ vanishes,
time averaging and spatial integration yield
\begin{equation}\label{rhow}
\frac{1}{\mathcal S}\int_\Omega\overline{\rho w}\,d\Omega
=\frac{1}{\mathcal S}\int_\Omega z(\overline C-\overline E)\,d\Omega.
\end{equation}

The time-mean gaseous-mass balance in a statistically
steady state also gives
\begin{equation}\label{Papp}
P_c=\frac{1}{\mathcal S}
\int_\Omega\overline E\,d\Omega,
\end{equation}
where $P_c$ is defined in Eq.~\eqref{Pc}.

The corresponding mean
condensation and evaporation heights are
\begin{equation}\label{zcapp}
\begin{aligned}
z_c&=\frac{1}{\mathcal S P_c}\int_\Omega z\overline C\,d\Omega,\\
z_e&=\frac{1}{\mathcal S P_c}\int_\Omega z\overline E\,d\Omega.
\end{aligned}
\end{equation}
Substituting Eq.~\eqref{rhow} into Eq.~\eqref{WPapp1} gives
$W_P=gP_c(z_c-z_e)$, as in Eq.~\eqref{WP}.

\section{Diagnosing $W_K$ from MERRA-2}
\label{app:WK}

We diagnose kinetic energy generation by horizontal pressure gradients from
three-hourly MERRA-2 pressure-level data. The calculation uses the horizontal
wind components $u$ and $v$, the height $Z$ of each isobaric surface, surface
pressure $p_s$, and the pressure-level coordinate $p$. The geopotential of an
isobaric surface is
\begin{equation}
\label{Z}
\Phi(p,\varphi,\lambda,t) = g Z(p,\varphi,\lambda,t).
\end{equation}
We use $g=9.80665~\mathrm{m\,s^{-2}}$. In pressure coordinates, the horizontal
pressure-gradient force per unit mass is $-\nabla_p \Phi$. Thus the local rate
of kinetic energy generation per unit mass is
\begin{equation}
c_K(p,\varphi,\lambda,t)
=
-\mathbf{u}\cdot\nabla_p\Phi
=
-\left(
u\frac{\partial \Phi}{\partial x}
+
v\frac{\partial \Phi}{\partial y}
\right),
\label{eq:ck_local}
\end{equation}
with units $\mathrm{W\,kg^{-1}}$.
The local rate in Eq.~\eqref{eq:ck_local} is the
mass-specific counterpart of the pressure-work density
$-\mathbf{u}\cdot\nabla_h p$ integrated in Eq.~\eqref{WK}.

The horizontal derivatives in Eq.~\eqref{eq:ck_local} were evaluated by
centered finite differences on each pressure surface, retaining a grid
point only when its centre and four immediate horizontal neighbours
have valid $u$, $v$ and $Z$ and lie above the local surface.
For longitude, periodic
boundary conditions were used:
\begin{equation}
\frac{\partial \Phi}{\partial x}
\simeq
\frac{1}{R_E\cos\varphi}
\frac{\Phi(p,\varphi,\lambda+\Delta\lambda)
      -\Phi(p,\varphi,\lambda-\Delta\lambda)}
     {2\Delta\lambda}.
\label{eq:dphidx}
\end{equation}
For latitude,
\begin{equation}
\frac{\partial \Phi}{\partial y}
\simeq
\frac{\Phi(p,\varphi+\Delta\varphi,\lambda)
      -\Phi(p,\varphi-\Delta\varphi,\lambda)}
     {2R_E\Delta\varphi}.
\label{eq:dphidy}
\end{equation}
Here $R_E=6371~\mathrm{km}$ is Earth's radius, and angular increments are in
radians. The two polar rows are excluded from the $c_K$ calculation, avoiding
the singular zonal metric factor.

Values were treated as invalid if non-numeric, missing-valued or outside
the permissive source-code screening bounds ($-500$ to $500$~m~s$^{-1}$
for wind components, $-1000$ to $100000$~m for $Z$, and
$10000$ to $110000$~Pa for surface pressure). Values greater than
$10^{14}$ in absolute magnitude, including the MERRA-2 fill flag,
were also rejected. The full five-point stencil and the constraint
$p\le p_s$ were applied at every valid pressure surface.
Invalid stencils were excluded from the resolved layer integral;
in the global area mean they contribute zero to a fixed-area
denominator, without being regarded as observed zero generation.
Separate validity fractions accompany the local zonal means.

At each horizontal grid point, the resolved pressure-column integral was
computed by trapezoidal integration over adjacent valid pressure levels:
\begin{equation}
K_0(\varphi,\lambda,t)
=
\sum_{k=1}^{N_p-1}
\frac{1}{2}
\left(c_{K,k}+c_{K,k+1}\right)
\frac{\Delta p_k}{g},
\label{eq:column_base}
\end{equation}
where $\Delta p_k=|p_{k+1}-p_k|$. Since $\Delta p/g$ is the atmospheric mass
per unit area between two pressure surfaces, $K_0$ has units
$\mathrm{W\,m^{-2}}$. The pressure-layer contributions are retained separately for the
study period and its constituent March--February intervals; their
split at 500~hPa uses each layer's pressure midpoint.

Because the lowest valid pressure level does not always coincide with the
local surface pressure, we computed three surface-layer corrections. Let $p_b$
be the lowest valid pressure level in a column, $c_b$ the corresponding value
of $c_K$, and $\Delta p_s=p_s-p_b$. If $\Delta p_s>0$, the correction used for
the main estimate assumes that $\mathbf{u}\cdot\nabla p=0$ at the surface and
that the mean local kinetic energy generation between $p_b$ and $p_s$ is
one half of its value at $p_b$:
\begin{equation}
K_{\mathrm{corr}}
=
K_0
+
\frac{1}{2}c_b\frac{\Delta p_s}{g}.
\label{eq:surface_corr}
\end{equation}
This is the value reported in the main text.
For the constant-value surface-layer treatment shown in Fig.~\ref{fig:WK}a,
we hold $c_K=c_b$ throughout the unresolved layer, giving
\[
K_{\mathrm{const}} = K_0 + c_b\frac{\Delta p_s}{g}.
\]

As a sensitivity estimate, we also extrapolated $c_K$ to the surface
from the lowest valid level and its \emph{immediately adjacent} valid
upper level (the archived \texttt{adjacent} mode). If $p_u$ is that
upper pressure level, with value $c_u$, then
\begin{equation}
c_s
=
c_b
+
(c_b-c_u)
\frac{p_s-p_b}{p_b-p_u},
\label{eq:surface_extrap_ck}
\end{equation}
and
\begin{equation}
K_{\mathrm{extr}}
=
K_0
+
\frac{1}{2}
(c_b+c_s)
\frac{\Delta p_s}{g}.
\label{eq:surface_extrap}
\end{equation}
If the immediately adjacent upper level was invalid, the extrapolated
correction was replaced by Eq.~\eqref{eq:surface_corr}, without
interpolation across the missing level.

Finally, global means were obtained by area-weighting the column values:
\begin{equation}
\overline{K}(t)
=
\frac{\sum_{i,j} K_{ij}(t) A_{ij}}
     {\sum_{i,j} A_{ij}},
\label{eq:global_mean}
\end{equation}
where the grid-cell area was computed from latitude-band boundaries,
\begin{equation}
A_{ij}
=
R_E^2\Delta\lambda
\left[
\sin\left(\varphi_{i+1/2}\right)
-
\sin\left(\varphi_{i-1/2}\right)
\right].
\end{equation}
The four estimates compared in Fig.~\ref{fig:WK}a are the resolved pressure-level
integral $\overline{K}_0$, the half-bottom-value surface correction
$\overline{K}_{\mathrm{corr}}$, the constant-value surface correction
$\overline{K}_{\mathrm{const}}$, and the extrapolated surface
correction $\overline{K}_{\mathrm{extr}}$.
The value discussed in the paper and shown in Fig.~\ref{fig:WvW} is
\[
W_K = \overline{K}_{\mathrm{corr}}.
\]
The differences between the adopted estimate and the constant-value
and extrapolated estimates characterize the sensitivity of $W_K$
to the unresolved surface layer  (Fig.~\ref{fig:WK}a).

\section{A Lorenz-type formulation for mass-related pressure work}
\label{app:Lorenz}

Here we develop a simplified mass-related
available-energy framework to examine the
relationship between surface-pressure-gradient
work and the temperature-related contribution
to kinetic energy generation
(Section~\ref{sec:discussion-why}).

Consider a horizontally closed, hydrostatic atmosphere with the column gaseous mass $m=p_s/g$,
where $p_s$ is surface air pressure. 
Writing $\hat p_s\equiv\mathcal S^{-1}\int_{\mathcal S}p_s\,dS$,
the mass-related available-energy functional is
\begin{equation}\label{eq:Am}
A_m=\frac{RT_0}{Mg\mathcal S}\int_{\mathcal S}
p_s\ln\frac{p_s}{\hat p_s}\,dS.
\end{equation}
This functional is nonnegative, vanishing only for horizontally
uniform surface pressure\footnote{The isothermal work equivalent of a
nonuniform pressure distribution was derived by
\citet[pp.~506--507]{Margules1901} in a form identical to
$\mathcal S A_m$ when the reference pressure is $\hat p_s$.}.

The local source of gaseous mass is $E-C$, using the net evaporation
and condensation rates defined in Section~\ref{sec:CCTc}. 
Define the column horizontal
gaseous-mass flux and the column net source as
\begin{equation}\label{eq:mass-defs}
\mathbf J=\int_0^\infty\rho\mathbf u\,dz,\qquad
q=\int_0^\infty(E-C)\,dz.
\end{equation}
Integrating the continuity
equation~\eqref{Ssource} over height gives
\begin{equation}\label{eq:masscontinuity}
\frac{1}{g}\frac{\partial p_s}{\partial t}
+\nabla_h\cdot\mathbf J=q.
\end{equation}

Differentiating Eq.~\eqref{eq:Am}, using
Eq.~\eqref{eq:masscontinuity}, and integrating by parts over the
closed horizontal domain yields
\begin{equation}\label{eq:massbud}
\frac{dA_m}{dt}=G_m-W_m,
\end{equation}
where
\begin{equation}\label{eq:Gm-isothermal}
G_m=\frac{RT_0}{M\mathcal S}\int_{\mathcal S}
q\ln\frac{p_s}{\hat p_s}\,d\mathcal S
\end{equation}
is the mass-related available-energy generation, and
\begin{equation}\label{eq:wk-isothermal}
W_m=-\frac{RT_0}{M\mathcal S}\int_{\mathcal S}
\mathbf J\cdot\nabla_h\ln p_s\,dS
\end{equation}
is the area-mean surface pressure-gradient work. 
In an isothermal atmosphere with constant molar mass $M$ and $T = T_0$,
we have $\nabla_h\ln p(z)=\nabla_h\ln p_s$. In this case,
$W_m$ equals the kinetic energy generation $W_K$
defined by Eq.~\eqref{WK}.

In steady state, $G_m=W_m$.
Equation~\eqref{eq:Gm-isothermal} shows that $G_m$
depends on the spatial covariance between the column
gaseous-mass source $q$ and surface pressure $p_s$.
For example, in the tropics, zonal-mean evaporation
varies relatively smoothly with latitude, whereas
precipitation exhibits pronounced maxima
\citep[Fig.~1a]{siler2021}.
More generally, evaporation is relatively distributed
in space and time, while intense condensation is
concentrated in precipitation systems, often associated
with lower surface pressure.
Such a distribution of gaseous-mass sources and sinks
can produce a positive covariance between $q$ and $p_s$.

For an order-of-magnitude estimate, we assume a
characteristic pressure difference
$\Delta p_s=10$~hPa between gaseous-mass source and
sink regions. Taking the mean tropical precipitation
rate $P\simeq3.6$~mm~day$^{-1}$ from GPCP V3.3
(Fig.~\ref{fig:WP}a), $\hat p_s=1000$~hPa,
and $T_0=300$~K, we obtain
\begin{equation}\label{eq:Gm-scale}
G_m\simeq P\frac{RT_0}{M}
\frac{\Delta p_s}{\hat p_s}
\simeq0.04~\mathrm{W\,m^{-2}}.
\end{equation}
The associated equivalent velocities for large-scale
net mass transport are of order $1$~cm~s$^{-1}$
or less \citep[e.g.,][p.~51]{lorenz67}.
Such estimates have contributed to the conventional
neglect of the dynamical importance of the
condensation-induced mass sink.

The small magnitude of $G_m$ does not imply that
pressure work at individual heights is small.
Consider the air-mass flow against
the surface-pressure gradient, which generates positive
kinetic energy near the surface. Mass continuity
requires a compensating return flow aloft (Fig.~\ref{fig:airflow-isobaric}).
We therefore separate the column gaseous-mass flux
into lower and upper contributions,
$\mathbf J=\mathbf J_l+\mathbf J_u$, where
\begin{equation}
\mathbf J_l=\int_0^{z_*}\rho\mathbf u\,dz,
\qquad
\mathbf J_u=\int_{z_*}^\infty\rho\mathbf u\,dz.
\end{equation}
Here $z_*$ denotes the height separating the opposing
mass flows, with $\mathbf J_l$ directed against
the surface-pressure gradient and $\mathbf J_u$
in the opposite direction. 
This idealized separation need not coincide with
the 500-hPa boundary used for the observational
diagnostics in Section~\ref{sec:resultsWK}.

Their contributions to the surface-pressure-gradient
work are
\begin{equation}
W_{m,l}+W_{m,u}=W_m=G_m,
\end{equation}
where
\begin{equation}
W_{m,l(u)}=-\frac{RT_0}{M\mathcal S}
\int_{\mathcal S}
\mathbf J_{l(u)}\cdot\nabla_h\ln p_s\,dS.
\end{equation}

Since the upper atmosphere is generally drier than
the lower atmosphere, the return flow carries less
water vapor. Denoting the difference in water-vapor
mass mixing ratio between the lower and upper
flows by $\Delta r=r_l-r_u$, and their characteristic
air-mass transport rate per unit area by $F$,
moisture conservation gives $P\simeq F \Delta r$.
For a representative surface-pressure difference
$\Delta p_s$, the positive lower-atmospheric work is
therefore
\begin{equation}\label{eq:Wm-lower}
W_{m,l}\simeq F\frac{RT_0}{M}
\frac{\Delta p_s}{\hat p_s}
\simeq\frac{G_m}{\Delta r}.
\end{equation}
For $\Delta r\simeq0.01$--$0.02$, the positive
lower-atmospheric pressure work can exceed $G_m$
by a factor of $50$--$100$.
With $G_m\simeq0.04$~W~m$^{-2}$,
this yields $W_{m,l}\simeq2$--$4$~W~m$^{-2}$,
with an approximately compensating negative
contribution $W_{m,u}$ from the upper atmosphere.

We now impose, as a dynamical constraint on our idealized
circulation, that the net upper-atmospheric kinetic energy
generation be negligible, $W_{K,u}\simeq0$.

We consider small horizontal temperature and surface
pressure differences about an isothermal reference
at temperature $T_0$.
Assuming that the horizontal temperature difference
is approximately independent of height,
hydrostatic balance gives, to first order,
\begin{equation}\label{eq:pressure-gradient-T}
\frac{RT}{M}\nabla_h\ln p(z)
\simeq
\frac{RT_0}{M}\nabla_h\ln p_s
+\frac{gz}{T_0}\nabla_h T.
\end{equation}
Thus, kinetic energy generation separates into
\begin{equation}\label{eq:WK-Wm-WT}
W_K\simeq W_m+W_T,
\end{equation}
where $W_m$, given by Eq.~\eqref{eq:wk-isothermal},
is the surface-pressure-related contribution, while
\begin{equation}\label{eq:WT}
W_T=-\frac{g}{T_0\mathcal S}
\int_{\mathcal S}
\mathbf J^{(z)}\cdot\nabla_h T\,dS,
\qquad
\mathbf J^{(z)}\equiv
\int_0^\infty z\rho\mathbf u\,dz
\end{equation}
is the temperature-related contribution.
While $W_m$ depends on the column-integrated
mass flux $\mathbf J$, $W_T$ depends on its
height-weighted counterpart $\mathbf J^{(z)}$.
Thus, the opposing lower and upper mass flows
can nearly cancel in the surface-pressure-related
work $W_m$, while their different heights allow a
substantial temperature-related contribution $W_T$.

Consider the upper return flow $\mathbf J_u$
defined above, with characteristic mass-flux-weighted
height $\bar z_u$.
Let $\Delta T>0$ denote the temperature difference
between the warmer low-pressure and cooler
high-pressure air columns.
The corresponding upper-atmospheric kinetic energy
generation is
\begin{equation}\label{eq:WKu-zero}
W_{K,u}\simeq W_{m,u}+W_{T,u}
\simeq F\left(
-\frac{RT_0}{M}\frac{\Delta p_s}{\hat p_s}
+\frac{g\bar z_u}{T_0}\Delta T
\right).
\end{equation}
The condition $W_{K,u}\simeq0$ therefore requires
\begin{equation}\label{eq:zDeltaT}
\frac{\Delta T}{T_0}
\simeq
\frac{H}{\bar z_u}
\frac{\Delta p_s}{\hat p_s},
\qquad
H\equiv\frac{RT_0}{Mg},
\end{equation}
where $H$ is the atmospheric scale height.
This means that the temperature difference required
to offset the negative contribution of a given
surface-pressure gradient aloft is not unique.
The higher the return flow, the smaller the required
temperature difference. For example, with
$T_0=300$~K and $\Delta p_s=10$~hPa
at $\hat p_s=1000$~hPa, an upper return flow
at the tropical tropopause height
$\bar z_u=16$~km requires a temperature difference
of only $\Delta T\simeq1.6$~K,
whereas for $\bar z_u=H\simeq8.8$~km,
$\Delta T\simeq3$~K is required.

When the positive lower-atmospheric work is
concentrated near the surface, its temperature-related
correction is comparatively small.
Consequently, $W_T\simeq W_{T,u}\simeq-W_{m,u}$ and
\begin{equation}\label{eq:WK-lower-final}
W_K \simeq W_{m,l} \simeq W_T.
\end{equation}

Combining Eqs.~\eqref{eq:Wm-lower} and
\eqref{eq:WK-lower-final}, and identifying the
lower-atmospheric kinetic energy generation with
the condensation-related expansion power $W_c$,
as proposed in Section~\ref{sec:discussion-why},
we obtain for the idealized circulation
\begin{equation}\label{eq:Wc-WT-Lorenz}
W_c\simeq W_{K,l}\simeq W_K\simeq W_T
\simeq\frac{G_m}{\Delta r}.
\end{equation}

This relation illustrates how the small mass-related
available-energy generation $G_m$ can correspond
to a much larger kinetic energy generation in the
lower atmosphere, while the same net work can be
expressed through the temperature-related
contribution $W_T$. The physical interpretation
of this correspondence is discussed in
Section~\ref{sec:discussion-why}.

\section*{Data and code availability}

GPM/DPR spectral latent-heating (SLH) Level-2 V07C
orbital data were obtained from NASA's Precipitation
Processing System (PPS) archive
(\url{https://arthurhouhttps.pps.eosdis.nasa.gov/gpmallversions/V07/}).
MERRA-2 three-hourly pressure-level data
(M2I3NPASM, version 5.12.4) were obtained from
NASA's Goddard Earth Sciences Data and Information
Services Center (GES DISC)
(\url{https://goldsmr5.gesdisc.eosdis.nasa.gov/data/MERRA2/M2I3NPASM.5.12.4/}). 
The processing code, definitions of archived variables,
computational procedures, intermediate results,
and reproducibility checks are available
in the accompanying Zenodo archive
\url{https://doi.org/10.5281/zenodo.24269780}.

\section*{Funding statement}

This work received institutional support from the
Biotic Pump Greening Institute and the
Institute for Advanced Study, Technical University
of Munich. No specific grant funding was received.

\section*{Conflict of interest statement}

The authors declare no conflict of interest.

\section*{Acknowledgments}

The authors have no additional acknowledgments to report.

\end{document}